\documentclass[preprint,showpacs,amsmath,amssymb]{revtex4}

\draft
\usepackage{graphicx,color}

\begin{document}

\title{ Violation of non-interacting $\cal V$-representability of  
the exact solutions of the Schr\"odinger equation 
for  a two-electron quantum dot in a homogeneous magnetic field}
\author{ M.Taut, P.Machon, and H.Eschrig }
\affiliation{ Leibniz Institute for Solid State and Materials Research, 
IFW Dresden\\
POB 270116, 01171 Dresden, Germany,\\
email: m.taut@ifw-dresden.de}

\date{\today}

\begin{abstract}
We have shown by using the exact solutions for the two-electron system
in a parabolic confinement and a homogeneous magnetic field
\cite{Taut-2einB, Taut-3einB} that both exact densities 
(charge- and the paramagnetic current density)
can be  non-interacting $\cal V$-representable (NIVR) only in a few special cases,
 or equivalently,
that an exact Kohn-Sham (KS) system does not always exist.
All those states at non-zero $B$ can be NIVR, which are 
 continuously connected to the singlet or triplet ground  states at $B=0$.
In more detail, 
for {\em singlets} (total orbital angular momentum $M_L$ is even)
both densities can be NIVR if the vorticity
 $\mbox{\boldmath$\gamma$}({\bf r})=\mbox{\boldmath $\nabla$}
\times \Big ({\bf j}^p({\bf r}) / n({\bf r}) \Big) $
of the exact solution vanishes.
For $M_L=0$ this is trivially guaranteed
 because the paramagnetic current density vanishes.
The vorticity based on the exact solutions
for the higher $|M_L|$ does not vanish, in particular for small r.
In the limit $r\rightarrow 0$ this can even be shown analytically.
For {\em triplets} ($M_L$ is odd) and if we 
assume circular symmetry for the KS system
(the same symmetry as the real system)
then only the exact states with $|M_L|= 1$ can be NIVR
 with KS states having angular momenta $m_1=0$ and $|m_2|=1$. 
Without specification of the symmetry of the KS system 
the condition for NIVR is that the small-r-exponents of the KS states 
are 0 and 1.

\end{abstract}


\maketitle

\large
\section{Introduction}
Semi-relativistic Current-Density Functional Theory (CDFT)
\cite{Vignale-Rasolt1,Vignale-Rasolt2}
has become one on the standard tools for the calculation 
of electronic ground state (GS) properties  in magnetic fields.
Apart from the practical issue of finding
an accurate and manageable energy functional,
the basic fact of {\em  non-interacting $\cal V$-representability} 
(NIVR) of the
exact electron density $n({\bf r})$ and the paramagnetic
current density $j^p({\bf r})$
is a prerequisite for the existence of a Kohn-Sham scheme. 
In a first step, we are investigating if 
 both {\em exact} densities derived from the correlated two-particle state 
can be represented by the same set of one-particle orbitals
\begin{equation}
\varphi_k({\bf r})=R_k({\bf r}) \; e^{i\zeta_k({\bf r})}
\label{one-e-state}
\end{equation}
in the form 
(atomic units $\hbar=m=e=1$ are used throughout)
\begin{eqnarray}
n({\bf r})&=&\sum_k^{occ} \varphi_k^*({\bf r})\; \varphi_k({\bf r})=
\sum_k^{occ} R_k^2({\bf r})
\label{density}
\\
{\bf j}^p({\bf r})&=& \frac{1}{2i} \sum_k^{occ} \Big[ \varphi_k^*({\bf r})\;
\mbox{\boldmath $\nabla$} \varphi_k({\bf r})-
\varphi_k({\bf r})\;\mbox{\boldmath $\nabla$}  \varphi_k^*({\bf r}) \Big]
= \sum_k^{occ} R_k^2({\bf r}) \; \mbox{\boldmath $\nabla$} \zeta_k({\bf r})
\label{current}
\end{eqnarray}
where the modulus $ R_k({\bf r}) \ge 0$ and the phase $\zeta_k({\bf r})$ 
of the Kohn-Sham wave-functions (KS-WF) are real functions.
{\em If} the KS-WFs are well defined, then the effective potentials  
${\cal V}^{eff}=(v^{eff}\; \mbox{and}\; {\bf A}^{eff})$ could 
be obtained from the KS equations 
as exercised in Ref.\cite{Wensauer-Roessler1}.
In this paper we are only discussing if and when KS-WFs exist.
Unlike in Density Functional Theory (DFT), NIVR in CDFT 
as defined above is neither guaranteed for nor restricted to GSs.

For vanishing magnetic field, 
DFT applies, which rests on some unique mappings. 
If $\Psi$ describes a many-body GS-WF
and $n({\bf r})$ and $v^{ext}({\bf r})$ the corresponding density and 
external potential (modulo a constant), then  of course 
$\Psi \stackrel {\cal D}{\rightarrow}  n({\bf r})$
 is unique, but also
$\Psi \stackrel {\cal C} {\rightarrow} v^{ext}({\bf r})$ ,
if infinitely high potential walls are excluded. 
Hohenberg and Kohn \cite{HK} proved that the mapping between $\Psi$ 
and $n({\bf r})$ is even one-to-one, and hence 
$n({\bf r}) \; \stackrel {{\cal C D}^{-1}} {\longrightarrow} v^{ext}({\bf r})$
is also unique
\begin{equation}
v^{ext}({\bf r}) \;\;
 \stackrel{\cal C}{\leftarrow}
\; \Psi \;\;
\begin{array}{c}
{\scriptstyle \cal D} \\[-2mm]
\rightleftharpoons\\[-2mm]
{\scriptstyle {\cal D}^{-1}}
\end{array}
\;\;n({\bf r})
\label{map}
\end{equation}
The situation of (\ref{map}) holds for both the interacting and the 
non-interacting cases. For a NIVR density $n({\bf r})$, the mapping 
${\cal CD}^{-1}$ for the non-interacting case yields the (effective) 
KS potential of the interacting system.
It has been shown (e.g. in \cite{Levy}) that not every 
mathematically well behaved density $n({\bf r})$ is GS density to some 
potential $v^{ext}({\bf r})$. Therefore, DFT has been based on 
functionals which are also defined for non-$v$-representable densities. 
Nevertheless, a KS potential (derivative of the density functional) can only exist for NIVR densities.

In the presence of a magnetic field and for (semi-relativistic)
 Current Density Functional Theory (CDFT),
the generalization of ${\cal D}^{-1}$ for the ground state 
still exists, but,
Vignale and Rasolt \cite{Vignale-Rasolt1, Vignale-Rasolt2}
just {\em presupposed} the existence of the generalization of $\cal C$ 
 \cite{Capelle-Vignale1} implying that NIVR and the existence 
of a KS scheme  has not been proven. 
Capelle and Vignale \cite{Capelle-Vignale1}, on the other hand, 
have shown that there can be several external potentials ${\cal V}^{ext}$ 
which provide the same wave functions and densities 
\begin{equation}
\begin{array}{c}{\cal V}^{ext}_1({\bf r})\\{\cal V}^{ext}_2({\bf r})\\ \cdots \end{array} \;\;
\begin{array}{c} \searrow\\ \rightarrow\\ \nearrow \end{array}\;\;
{\bf \Psi} 
\rightleftharpoons
{\cal N}({\bf r})
\label{CDFT-maps}
\end{equation}
where ${\cal V}^{ext}({\bf r})$ and ${\cal N}({\bf r})$ represent 
 both external potentials ($v^{ext}({\bf r})$ 
and $\bf A^{ext}({\bf r})$)
and both densities ($n({\bf r})$ and ${\bf j}^p({\bf r})$), respectively.
Hence, $\cal C$ cannot exist anymore as a unique mapping. 
For Spin Density Functional Theory (SDFT) the same problem 
was first pointed out by von Barth and Hedin \cite{Barth-Hedin}, 
and later analyzed in detail in 
\cite{Eschrig-Pickett1,Capelle-Vignale2}.
For our model system the l.h.s. of (\ref{CDFT-maps})
 is obvious from the fact that 
the exact densities ${\cal N}({\bf r})$ 
are determined by the effective frequency 
\begin{equation}
{\widetilde \omega}=\sqrt{\omega_0^2+(\omega_c/2)^2}
\end{equation}
alone and not by the external confinement frequency $\omega_0$ 
and the cyclotron frequency $\omega_c=B/c$ independently (see Sect.II).
In other words, all combinations of $\omega_0$ and $\omega_c$, 
which provide the same $\widetilde \omega$, provide the same densities. 
This fact rules the existence of the mapping $\cal C$ out, but 
does neither prove nor rule out NIVR or the existence of a KS system. 

Wensauer and R\"ossler \cite{Wensauer-Roessler1} used the scaling property
of our quantum dot model system 
\begin{equation}
H(\omega_0,\omega_c)=H(\omega_0\rightarrow{\widetilde \omega},
\omega_c\rightarrow 0)
+\frac{\omega_c}{2}\,L_z\;\;,
\label{scaling}
\end{equation}
where $L_z$ is the total orbital angular momentum, in order to 
apply the consequences from the  Hohenberg-Kohn theorem to non-zero fields.
Indeed, (\ref{scaling}) means that the Hamiltonian for a non-zero magnetic field 
$H(\omega_0,\omega_c)$ has the same eigen-functions as the Hamiltonian 
for zero magnetic field and 
 the effective confinement frequency  ${\widetilde \omega}$.
Only the eigenvalues are shifted by $(\omega_c / 2) M_L$, where 
$M_L$ is the total orbital angular momentum.
The point is, however, that this does {\em not} mean that 
the GS densities  ${\cal N}({\bf r})$ for all $\omega_c$ (for fixed 
$\omega_0$) are NIVR. Instead, this conclusion applies only to those cases, 
where the corresponding reference state at zero $\omega_c$ is a ground state. 
This are the states with $M_L=0$, which are special insofar, as 
the paramagnetic current vanishes even for non-zero magnetic field.
Their argument does not say anything about NIVR of the other cases.

In the present paper we investigate, under which conditions
 the exact GS densities ${\cal N}({\bf r})$ of our model system 
can be represented by KS orbitals. For making the conclusions 
gauge independent, we also matched the gauge independent {\em vorticity}
\begin{equation}
\mbox{\boldmath$\gamma$}({\bf r})=\mbox{\boldmath $\nabla$} 
\times \frac{{\bf j}^p({\bf r})} {n({\bf r})}
\label{def-vorticity}
\end{equation}
instead of the paramagnetic current density ${\bf j}^p({\bf r})$.
We will show that for a typical quantum dot model 
(two electrons in a parabolic confinement  and a magnetic field)
the exact GS densities are {\em not} generally NIVR. 
This is not a sophistry or subtleness, but 
the violation of NIVR is massive.
This suffices to 
disprove  the assumption of {\em general} NIVR and the {\em general}
existence of a KS system for the GSs.
On this background, all semi-relativistic 
CDFT calculations and  functionals  have to be considered with caution.
On the other hand, this does not mean that all results are utterly wrong. 
However, this fact undermines the credibility of semi-relativistic 
CDFT as a tool for making forecasts.

\section{Specification of the model and exact densities}
\subsection{Model Hamiltonian}
We consider a two-dimensional two-electron system 
(with Coulomb interaction between the electrons) in
a harmonic scalar potential $v^{ext}(r)=(1/2)\;\omega_0^2 \; r^2$ 
and a magnetic field perpendicular to the plane ${\bf B}=B \;{\bf e}_z$ 
represented by the vector potential 
(in symmetric gauge) 
${\bf A}^{ext}({\bf r})=(1/2) \;{\bf B}\times {\bf r}=
(1/2)\;B\;r\;{\bf e}_{\alpha}$. We introduced cylinder coordinates 
$(r,\alpha,z)$ with the cylinder axis perpendicular to the plane.
The Hamiltonian reads
\begin{equation}
H=\sum\limits^2_{i=1}\biggl\{{1\over 2}\biggl({\bf p}_i+
{1\over c}{\bf A}^{ext}({\bf r}_i)\biggl)^2 +
{1\over 2}\;\omega_0^2 \;r_i^2\biggl\}+{1 \over |{\bf r}_2-{\bf r}_1|}
\label{H}
\end{equation}
This is a widely used effective Hamiltonian
 model for a two-electron quantum dot.
The interaction of the spins with the magnetic field 
 $H_{spin}=g^*\; \sum\limits_{i=1}^3 \;{\bf s}_i \; \cdot{\bf B}$ 
is omitted (by chosing $g^*=0$) for two reasons.
First, in semi-relativistic 
effective Hamiltonian  theory $g^*$ is a material dependent 
parameter well below the vacuum value $g=2$. Therefore, the limiting 
case of vanishing $g^*$ ought to be covered by an exact theory. 
Second, the Zeeman term would make the Hamiltonian spin dependent 
and would necessitate a description of the system by the
 {\em spin} density $n_s({\bf r})$ 
(instead of the total density $n({\bf r}$))
and the paramagnetic current density ${\bf j}^p({\bf r})$ 
\cite{Vignale-Rasolt1}. The introduction of the spin density in
SDFT produces its own problems 
\cite{Eschrig-Pickett1,Eschrig-Pickett2,Capelle-Vignale2}
even for vanishing magnetic field, 
which we do not want to let interfere with the problems 
produced by the magnetic field.

\subsection{Exact solutions of the Schr\"odinger equation}
The Schr\"odinger equation with the Hamiltonian 
(\ref{H}) can be solved not only by reduction to the numerical 
solution of an (ordinary) radial Schr\"odinger equation \cite{Merkt},
but even analytically for a 
 discrete, but infinite set of effective
 frequencies $\widetilde{\omega}$ \cite{Taut-2einB, Taut-3einB}.
If we introduce relative and center of mass  coordinate
\begin {equation}
{\bf r}={\bf r}_2-{\bf r}_1~~~~~;~~~~~{\bf R}={1\over 2}({\bf r}_1+{\bf
r}_2)
\end{equation}
the Hamiltonian (\ref{H}) decouples exactly.
\begin{equation}
H=2 \;H_r+{1\over 2} \;H_R+H_{spin}
\end{equation}
The Hamiltonian for the c.m. motion agrees with the 
Hamiltonian of a non-interacting quasi particle 
\begin {equation}
H_R={1\over 2}\biggl[{\bf P}+
{1\over c}{\bf A}_R\biggr]^2+{1\over
2}\omega_R^2\; R^2
\label{H_R}
\end{equation}
and only the relative Hamiltonian contains the 
electron-electron interaction 
\begin {equation}
H_r={1\over 2}\biggl[{\bf p}+
{1\over c}{\bf A}_r\biggr]^2+{1\over
2}\omega_r^2\;r^2+\frac{1}{2r}\;\; ,
\label{H_r}
\end{equation}
where we introduced rescaled parameters
$\omega_R= 2\omega_0$, ${\bf A}_R=2{\bf A}({\bf R})$,
$\omega_r={1\over 2}\omega_0$, ${\bf A}_r={1\over 2}{\bf A}({\bf r})$
(the index '$r$'  and '$R$' refers to the relative
and c.m. coordinate systems, respectively).
The decoupling of $H$ allows the ansatz 
\begin {equation}
\Phi=\xi({\bf R}) \; \varphi({\bf r})\; \chi(s_1,s_2)
\label{Phi}
\end{equation}
where $\chi(s_1,s_2)$ are the singlet or triplet spin eigen-functions.

The eigen-functions of the c.m. Hamiltonian (\ref{H_R})  
have the form
\begin{equation}
\xi={e^{iM\cal{A}}\over \sqrt{2\pi}}~~{U_M(R)\over R^{1/2}}=
\frac{e^{iM\cal{A}}}{\sqrt{2\pi}}~~R_M(R)
~~~~~;~~~~~M=0,\pm 1,\pm 2,\ldots
\label{xi_R}
\end{equation}
where the polar coordinates of the c.m. vector  are denoted by 
$(R,\cal{A})$ 
 and the radial functions $U_M(R)$ and $R_M(R)$ 
can be found in standard textbooks.

With the following ansatz for the relative motion 
\begin{equation}
\varphi={e^{im\alpha}\over \sqrt{2\pi}}~~{u_m(r)\over
r^{1/2}}~~~~~;~~~~~m=0,\pm 1,\pm 2,\ldots
\label{phi_r}
\end{equation}
the Schr\"odinger equation 
$H_r\,\varphi({\bf r})=\epsilon_r\,\varphi({\bf r})$
gives rise to a radial Schr\"odinger equation for $u(r)$
\begin{equation}
\biggl\{-{1\over 2}~{d^2\over dr^2}+{1\over 2}\biggl(m^2-{1\over 4}\biggr)
{1\over r^2}+{1\over 2}\;\widetilde\omega_r^2\; r^2+\frac{1}{2r}
\biggl\}u(r)=\widetilde \epsilon_r \; u(r)
\label{rad-SGl}
\end{equation}
where the polar coordinates for the relative
 vector are denoted by $(r,\alpha)$,
$\widetilde\omega_r={1\over 2}\widetilde\omega$,  
$\widetilde \epsilon_r=\epsilon_r-{1\over 4}\, m \,
\omega_c$, and
$\omega_c={B \over c}$.
The solutions are subject
to the normalization condition $\int\limits^\infty_o dr|u(r)|^2=1$.
The Pauli principle demands that (because of
 the particle exchange symmetry of the spin eigen-functions)
in the singlet and triplet state, the relative angular momentum 
$m$ has to be  even or odd, respectively.
There is no constraint for the c.m. angular momentum $M$
 following from Pauli principle.
Because of the orthogonality of the coordinate transformation, the 
above described solutions are eigen-functions of
 the total orbital angular momentum with the eigenvalue $M_L=M+m$.

\begin{figure}[htbp]
\centering
\includegraphics*[scale=0.5]{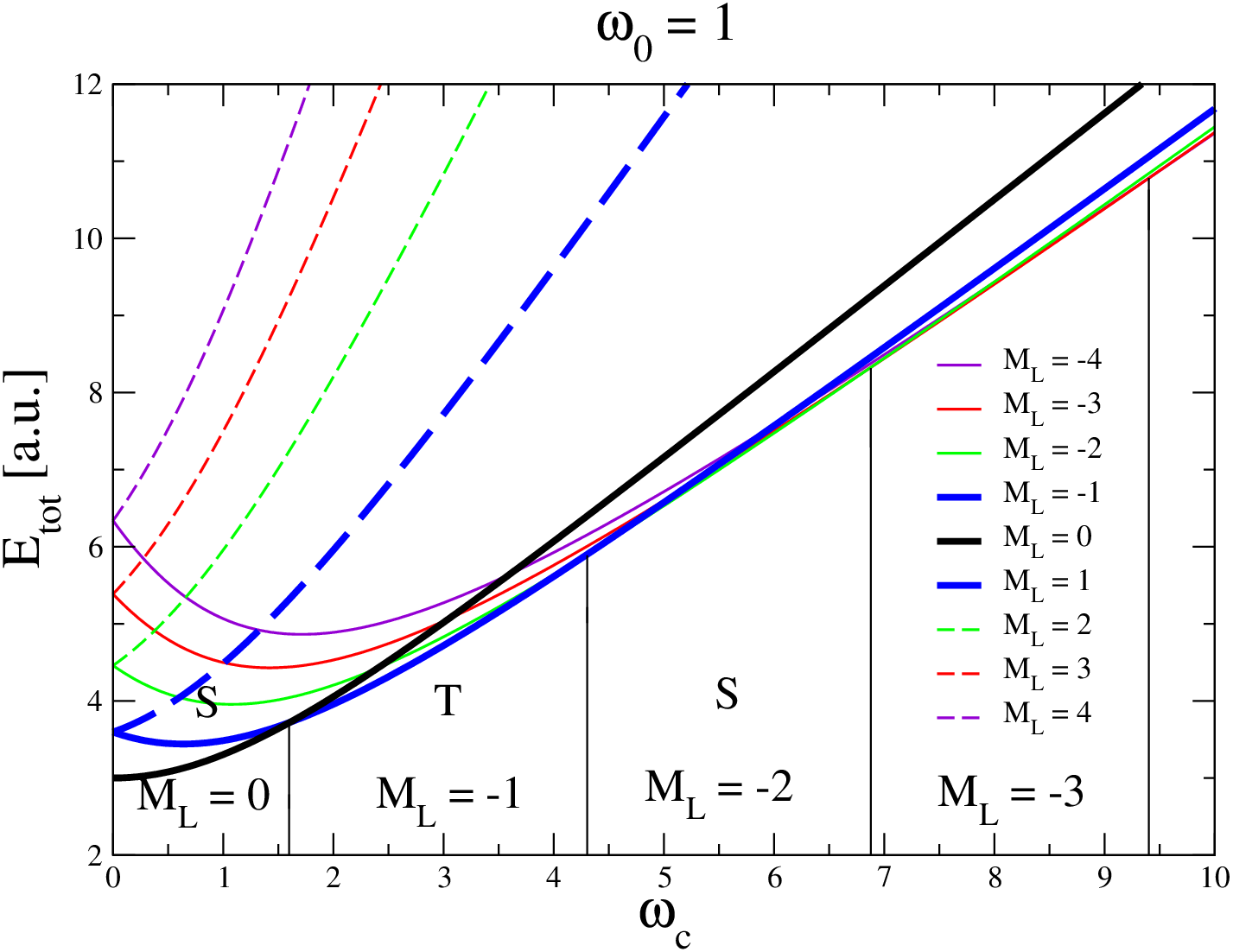}
\caption{(color online) 
Total energy for fixed confinement frequency $\omega_0=1$ 
versus cyclotron frequency $\omega_c$ (i.e. magnetic field).
The c.m. system is always in the ground state with $M=0$. The 
relative angular momentum $m$ is varied. The vertical lines 
show where the 
total orbital angular momentum $M_L=M+m$ of the ground state changes.
S and T indicates whether the ground state is singlet or triplet. 
Thick lines indicate states which can be NIVR.}
\label{fig-s-t}
\end{figure}

Fig.\ref{fig-s-t} shows that the modulus of the 
orbital angular momentum of the ground state grows stepwise
  with increasing 
magnetic field. This implies that the spin state oscillates 
between singlet and triplet \cite{Merkt2}.
(Quenching of the singlet state 
for higher magnetic fields due to a Zeeman term is not included in our model.)
States with  c.m. excitations are not included in the figure,
 because they are never ground states.

\subsection{Exact densities}

\begin{figure}[htbp]
\centering
\begin{minipage}[c]{18cm}
\includegraphics*[scale=0.29]{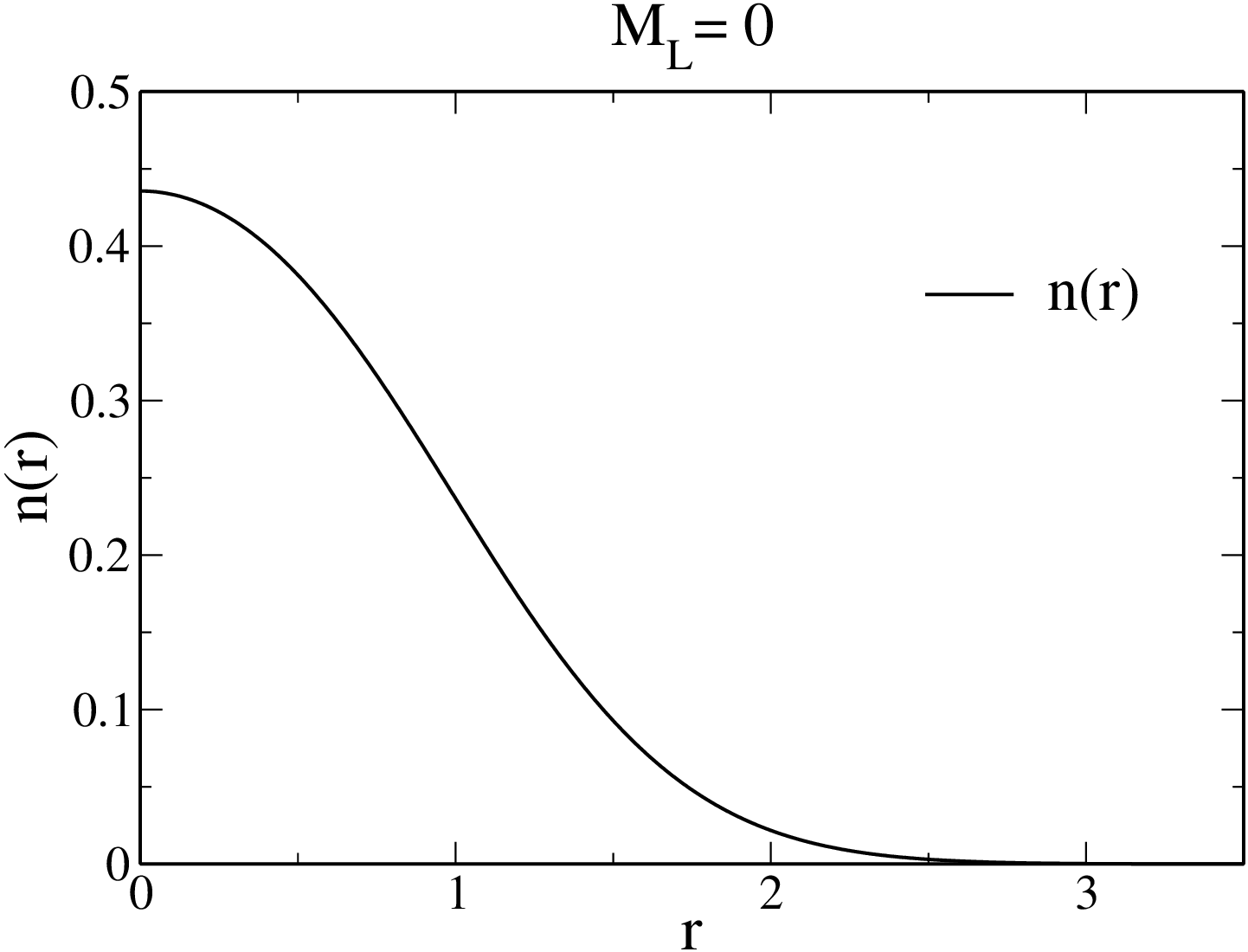}
\includegraphics*[scale=0.29]{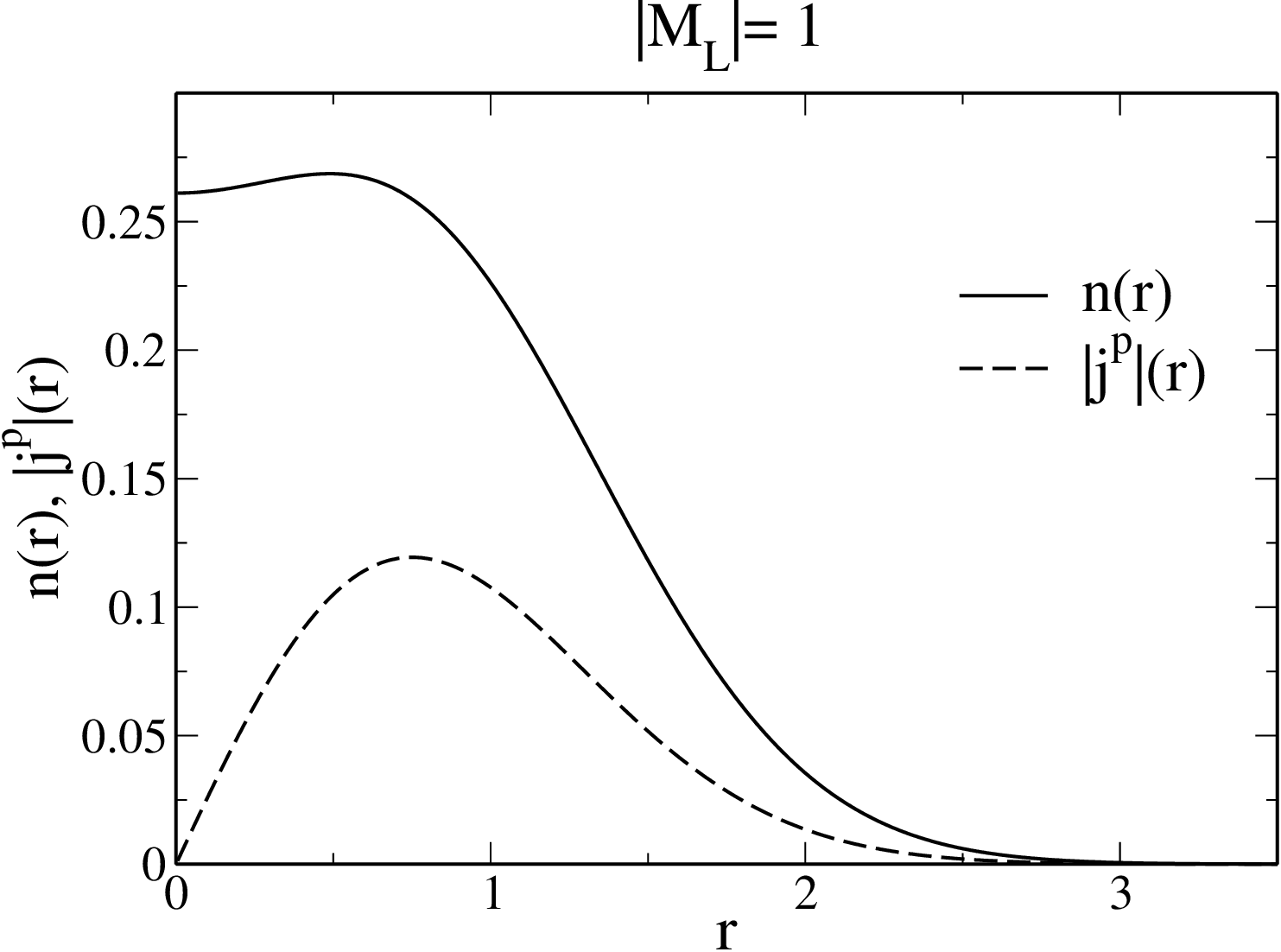}
\end{minipage}
\begin{minipage}[c]{18 cm}
\includegraphics*[scale=0.29]{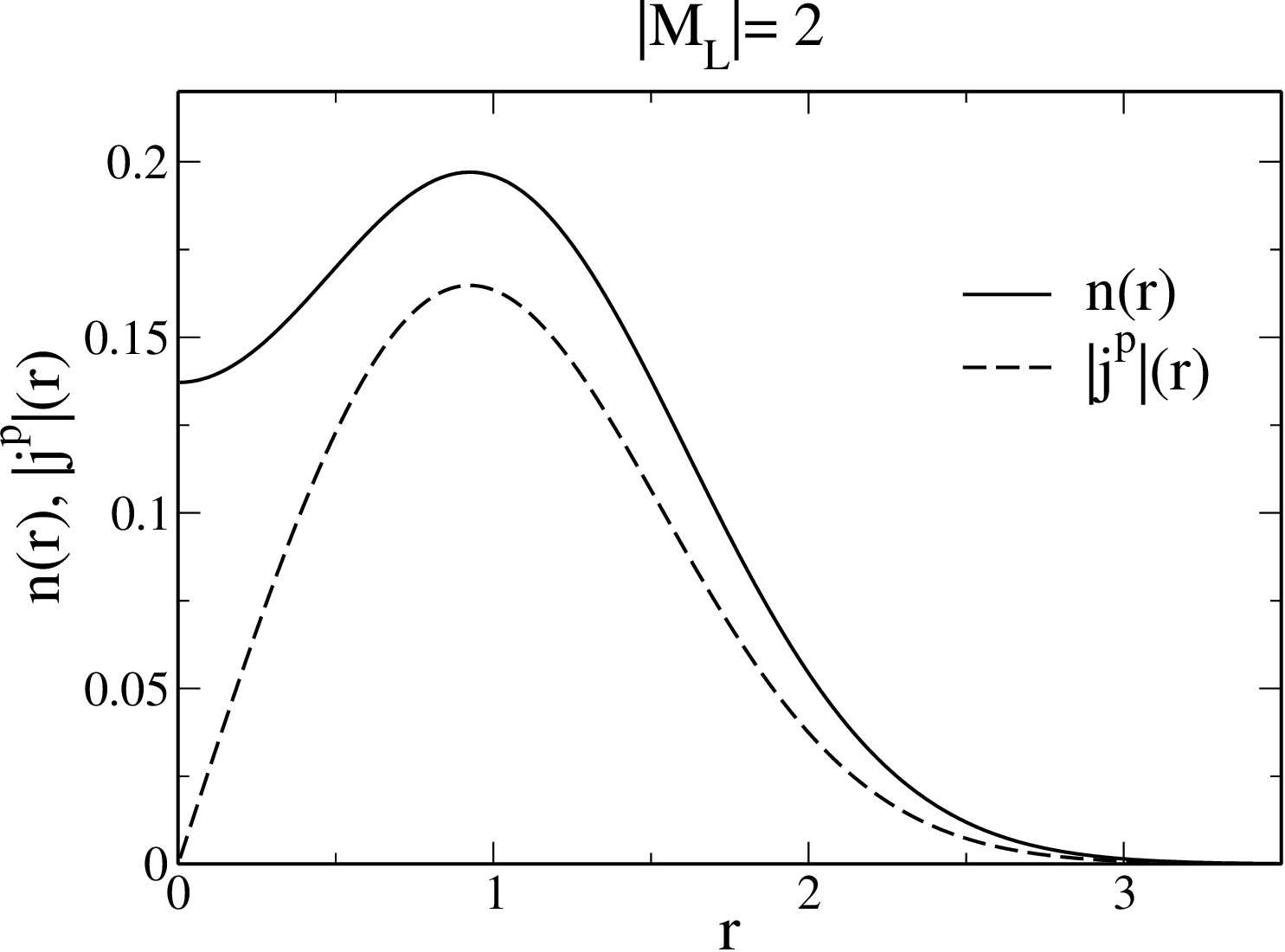}
\includegraphics*[scale=0.29]{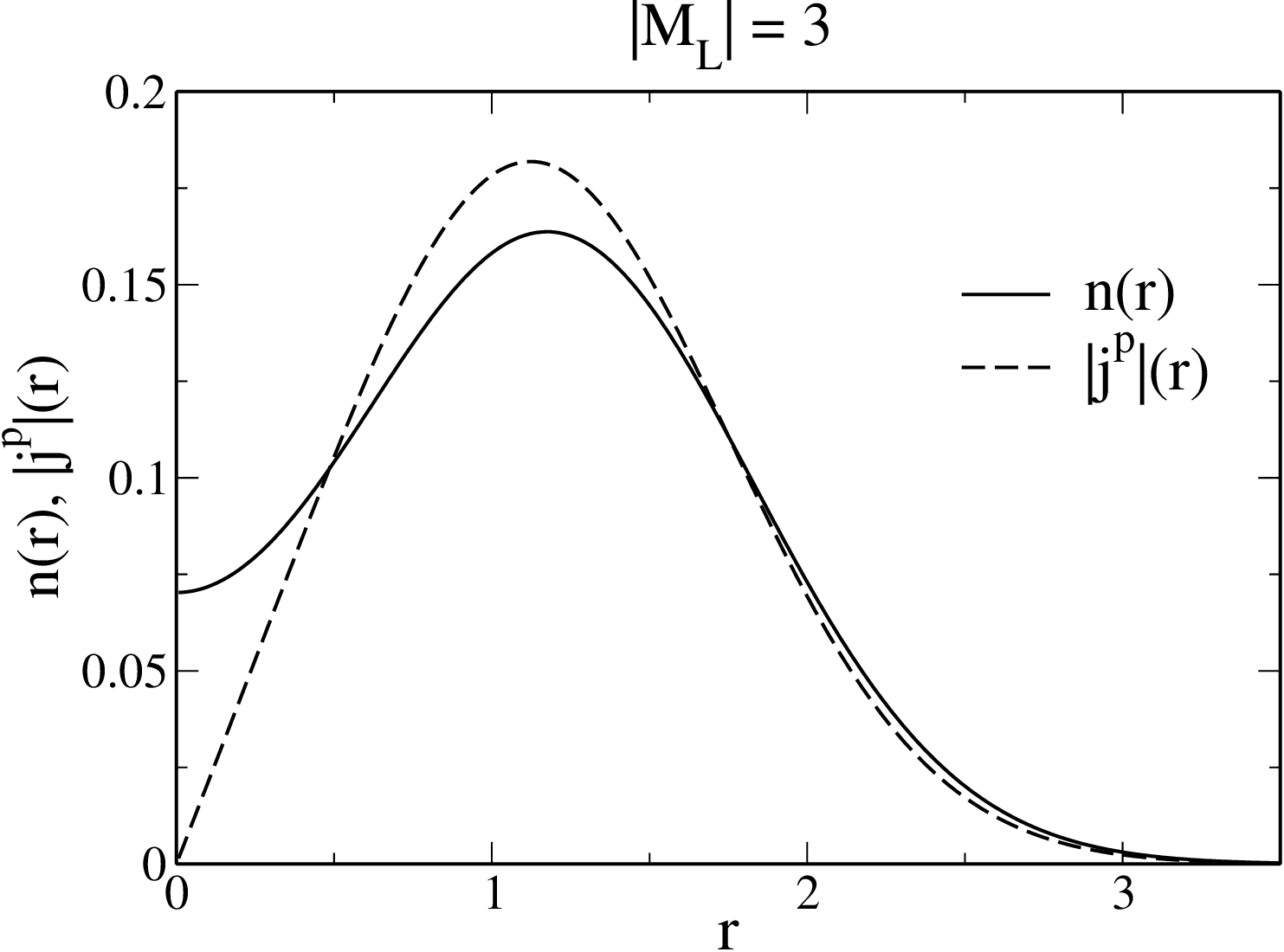}
\end{minipage}
\caption{
Exact densities and paramagnetic current densities (both in a.u.) 
for $\tilde{\omega}=1$
and the orbital angular momenta given
in the titles of the figures. The sign of $j^p(r)$ agrees 
with the sign of $M_L$.}
\label{fig-n-j}
\end{figure}

With (\ref{xi_R}) and  (\ref{phi_r}), we obtain for the total density 
\begin{equation}
n({\bf r})= 2 \int d{\bf r'}\; |\Phi({\bf r},{\bf r'})|^2
\end{equation}
the general expression
\begin{equation}
n(r)=\frac{1}{2 \pi^2} \int_0^{2\pi} d\alpha \int_0^\infty dr'\;
\bigg[ R_M\Big(\sqrt{r^2+\frac{1}{4}r'^2+r r' cos\alpha}\Big) \bigg]^2 \;
\Big[u_m(r') \Big]^2
\end{equation}
Because  we are interested in the ground state only, we can safely use 
the c.m. state for  $M=0$:
$R_0(R)=2\sqrt{\widetilde{\omega}} \;exp(-\widetilde{\omega} R^2)$ 
which allows to do one integration analytically leaving us with 
\begin{equation}
n(r)=\frac{4 \widetilde{\omega}}{\pi} e^{-2 \widetilde{\omega} \, r^2} 
\int_0^\infty dr'\; e^{-(\widetilde{\omega}/2) r'^2} \;
I_0(2 \widetilde{\omega} r r') \Big[u_m(r') \Big]^2
\label{n}
\end{equation}
where $I_n(x)$ are the modified Bessel functions.

The general expression for the paramagnetic current density
\begin{equation}
{\bf j}^p({\bf r})=-i \int d{\bf r'}\; 
\Big[
\Phi^*({\bf r},{\bf r'}) \mbox{\boldmath $\nabla$} \Phi({\bf r},{\bf r'}) -
\Phi({\bf r},{\bf r'}) \mbox{\boldmath $\nabla$} \Phi^*({\bf r},{\bf r'})
\Big]
\end{equation}
is somewhat complicated. Therefore  we give here only the formula for 
$M=0$
\begin{equation}
{\bf j}^p({\bf r})= {\bf e}_{\alpha}\; m \;\frac{4 \widetilde{\omega}}{\pi}
 e^{-2 \widetilde{\omega} \, r^2}
\int_0^\infty dr'\; e^{-(\widetilde{\omega}/2) r'^2} \;
\frac{I_1(2 \widetilde{\omega} r r')}{r'} \Big[u_m(r') \Big]^2
={\bf e}_{\alpha}\; j^p(r)
\label{j_p}
\end{equation}
As to be expected, the paramagnetic current density 
 points in azimuthal direction ${\bf e}_{\alpha}$, 
and the scalar $j^p(r)$  depends only on the distance $r$ from the center 
and not from the azimuthal angle .

Although both formulas (\ref{n}) and (\ref{j_p}) rely on the 
functions $u_m(r)$, which are  solutions 
of (\ref{rad-SGl}), the analytical behavior for $r \rightarrow 0$ 
can be expressed in terms of two positive definite integrals
\begin{eqnarray}
A_0&=&\int_0^\infty dr \; e^{-(\widetilde{\omega}/2) r^2} \;
 \Big[u_m(r) \Big]^2  \\
A_2&=&\int_0^\infty dr \; r^2\; e^{-(\widetilde{\omega}/2) r^2} \; 
 \Big[u_m(r) \Big]^2 
\end{eqnarray}
After power series expansion of $I_n(x)$, we obtain 
\begin{eqnarray}
n(r) &\rightarrow& \frac{4 \,\widetilde{\omega}}{\pi} 
e^{-2\, \widetilde{\omega} \, r^2} 
\Big[ A_0+A_2 \;\widetilde{\omega}^2 \;r^2 + \cdots \Big]
\label{n0}\\
j^p(r) &\rightarrow& m \frac{4 \,\widetilde{\omega}^2}{\pi}
e^{-2 \,\widetilde{\omega} \, r^2} \;  r 
\Big[ A_0+\frac{1}{2} A_2 \;\widetilde{\omega}^2\; r^2 + \cdots \Big]
\label{jp0}
\end{eqnarray}
For the origin this means that
$n(0)= 4 \,\widetilde{\omega}\,A_0/\pi$ is always finite and 
$j_p(0)=0$  always vanishes. 
On the other hand, 
the derivative of the density at the origin $\frac{d\, n}{dr}(0)=0$ 
vanishes, but of the paramagnetic current density 
$\frac{d\, j_p}{dr}(0)=m (4\, \widetilde{\omega}^2 /\pi) A_0$ is finite, 
unless $m=0$.
Besides, there is 
a relation which does not rely on the radial functions at all
\begin{equation}
\frac{d \,j_p}{dr}(0)=m\; \widetilde{\omega} \;n(0)
\end{equation}
The exact vorticity, which has the form  
$\mbox{\boldmath$\gamma$}({\bf r})={\bf e}_z\;\gamma(r)$, reads
in this limit
\begin{equation}
\gamma(r) \rightarrow m\,2\;\widetilde{\omega} \;\Big(1- \widetilde{\omega}^2 \;
\frac{A_2}{A_0} \; r^2 +\cdots \Big)
\label{gamma0}
\end{equation}
As will be seen below, the limit $r\rightarrow0$ is decisive for our proof.
The overall behavior of the densities is shown in Fig.\ref{fig-n-j}.

\section{Determine Kohn-Sham orbitals from exact density and paramagnetic 
current density}
Our consideration is in a sense  complementary to the approach in  
\cite{Wensauer-Roessler1}, where 
the triplet GS with $M_L=-1$ was considered.
We are going to point out that this is the only triplet GS  
 for which can be NIVR.

\subsection{Singlet state: }
In the KS system we assume that both electrons occupy the same  
orbital state of the form 
$\varphi({\bf r})=R({\bf r}) \; e^{i\zeta({\bf r})}$
 with different spins. We do not presuppose that this is an eigenstate 
of the orbital angular momentum, because there is no theorem which 
proves that orbital angular momentum is conserved in the KS system,
if it is conserved in the real system.
 Now we demand that both densities ($n, {\bf j}^p$) 
in the KS and in the real system agree.
\begin{equation}
n_{KS}({\bf r})=2 \;[R({\bf r})]^2  
\stackrel{!}{=} n_{exact}(r)
\label{det-n}
\end{equation}
\begin{equation}
{\bf j}^p_{KS}({\bf r})= 
2\; [R({\bf r})]^2 \; \mbox{\boldmath $\nabla$} \zeta({\bf r}) 
\stackrel{!}{=} j^p_{exact}(r) \; {\bf e}_{\alpha}
\label{det-j}
\end{equation}
Equation (\ref{det-n}) shows that the real part depends only on $r$ 
and reads
\begin{equation}
R(r)=\sqrt{\frac{1}{2}\; n_{exact}(r)}
\label{R-KS}
\end{equation}
Inserting the gradient in polar coordinates 
\begin{equation}
\mbox{\boldmath $\nabla$} \zeta(r,\alpha)=
{\bf e}_r\; \frac{\partial \zeta}{\partial r} +
{\bf e}_\alpha \;\frac{1}{r} \frac{\partial \zeta}{\partial \alpha}
\label{grad}
\end{equation}
into (\ref{det-j}) and using (\ref{det-n}) provides two equations for 
$\zeta(r, \alpha)$:
\begin{equation}
\frac{\partial \zeta(r,\alpha)}{\partial r}\stackrel{!}{=}0
 \;\; \rightarrow \;\; \zeta=\zeta(\alpha)
\label{zeta-KS-r}
\end{equation}
\begin{equation}
\frac{\partial \zeta (\alpha)}{\partial \alpha}\stackrel{!}{=}
r\frac{j^p_{exact}(r)}{n_{exact}(r)}
\equiv \mu_{exact}(r) \;\; \rightarrow 
\;\; \mu_{exact}(r)=const. 
\label{zeta-KS-alpha}
\end{equation}
If a function of $\alpha$ has to agree with a function of $r$ 
(left equation of (\ref{zeta-KS-alpha})), then both sides have to be a 
constant. Consequently, {\em for NIVR the function $\mu_{exact}(r)$ 
has to be constant}.

Before investigating this issue further, 
we introduce the vorticity (\ref{def-vorticity}) of the exact system.
By using the special form of the 
$curl$  of a vector field $\bf v$, which 
points in ${\bf e}_\alpha$ direction and  the modulus of which 
depends only on $r$,
\begin{equation}
\mbox{\boldmath $\nabla$}\times v(r) \,{\bf e}_\alpha=
 \frac{1}{r} \frac{d}{dr}\bigg ( r \,v(r) \bigg ) {\bf e}_z
\label{curl}
\end{equation}
we see that the vorticity  of the exact solutions 
has the general form
\begin{equation}
\mbox{\boldmath$\gamma$}_{exact}({\bf r})= \frac{1}{r}\frac{d}{dr} 
\; \bigg( r\frac{j^p_{exact}(r)}{n_{exact}(r)} \bigg)\; {\bf e}_z 
\equiv \gamma_{exact}(r) \; {\bf e}_z
\label{vorticity-exact}
\end{equation}
where ${\bf e}_z$ is the unit vector perpendicular to the 2D system,
and from (\ref{zeta-KS-alpha}) it follows
\begin{equation}
\gamma_{exact}(r)=\frac{1}{r}\frac{d}{dr}\; \mu_{exact}(r)
\stackrel{!}{=}0
\label{vorticity-z-exact}
\end{equation}
This means that  {\em for NIVR the exact 
vorticity $\gamma_{exact}(r) $ has to vanish}.

\begin{figure}[htbp]
\centering
\includegraphics*[scale=0.4]{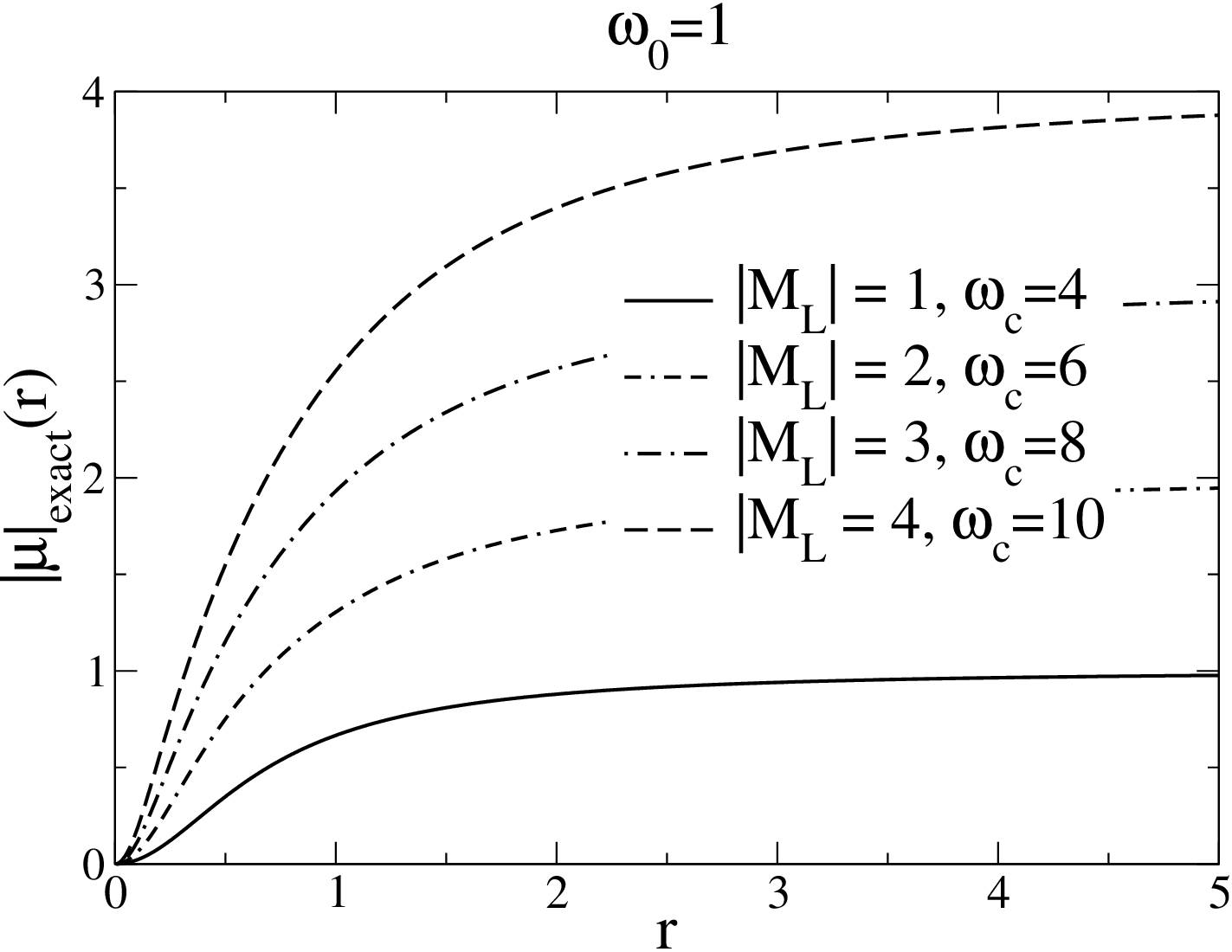}
\caption{$\mu_{exact}(r)$ (in a.u.) for $\omega_0=1$ and 
typical cyclotron frequencies $\omega_c$, 
where the state with negative $M_L$ is the ground state (see Fig.\ref{fig-s-t}).
The sign of $\mu_{exact}(r)$ agrees with the sign of $M_L$.
}
\label{fig-z}
\end{figure}

\begin{figure}[htbp]
\centering
\includegraphics*[scale=0.4]{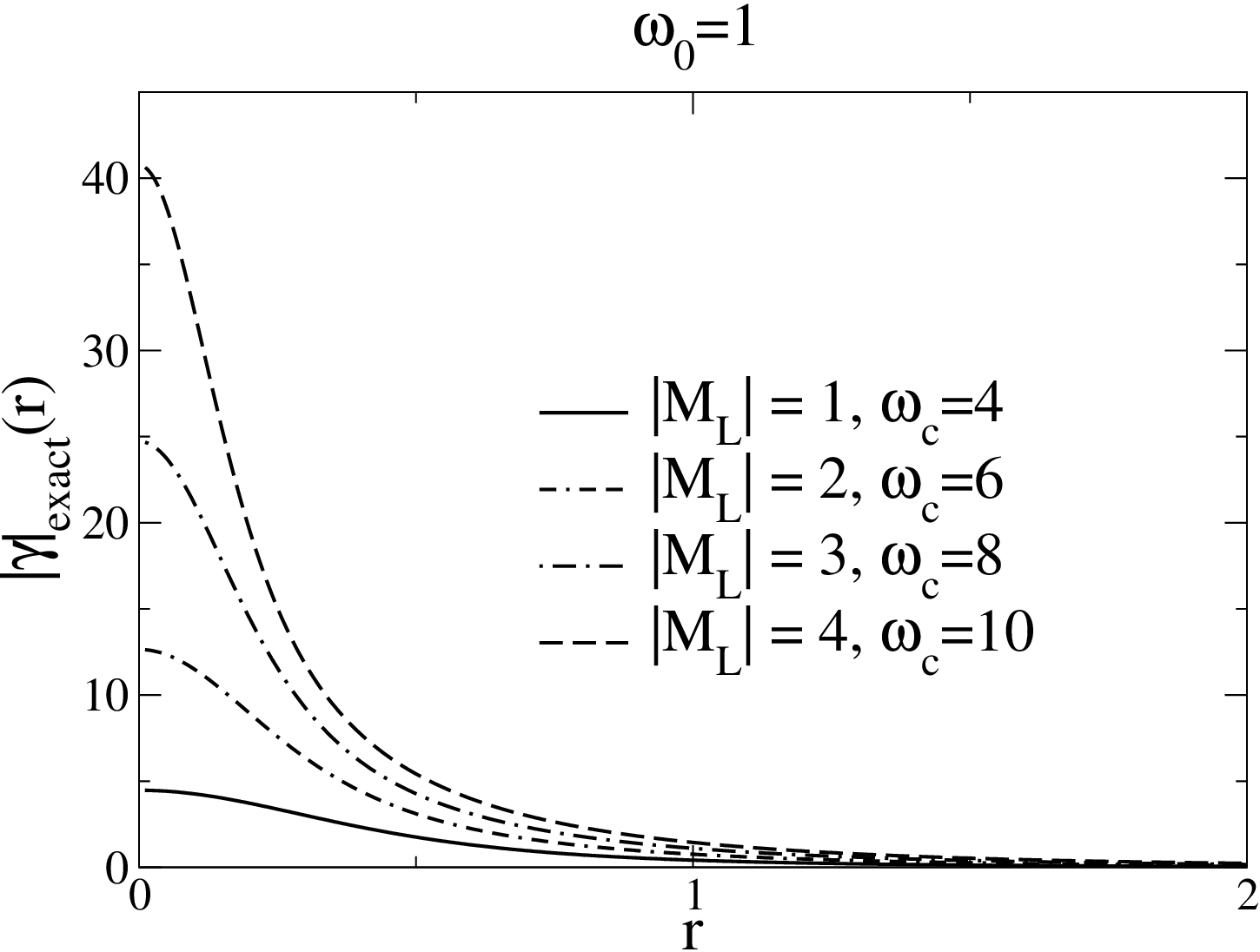}
\caption { Vorticities (in a.u.) for the same external fields as
Fig.\ref{fig-z}. 
The sign of $\gamma_{exact}(r)$ agrees with the sign of $M_L$.}
\label{fig-vort}
\end{figure}

The pure fact of violation of NIRV 
(without showing the quantitative extent in $r$-space) 
can already be seen {\em analytically} in the limit for
small $r$ given in (\ref{gamma0}), 
which provides $\gamma_{exact}(0)=2\, {\widetilde \omega}\, M_L$.
This value can be considered as a simple qualitative measure for the degree 
of violation. 
The numerical curves in Figs. \ref{fig-z} and \ref{fig-vort} show 
that the violation of NIVR in an extented small-$r$-region 
is massive. 
There is only one singlet state, where NIVR is trivially guaranteed 
for all $r$. 
This is $M_L=0$, where the paramagnetic current and consequently 
$\mu_{exact}(r)$ and $\gamma_{exact}(r)$ vanish exactly.
(As a side product we observe that $\mu_{exact}(r)$ converges for large $r$ 
to a constant, which agrees with the
 (total) orbital angular momentum $M_L$ 
of the exact state.)

Next we are investigating the vorticity of the KS system.
From (\ref{def-vorticity}) and               
(\ref{det-n},\ref{det-j}) it follows that the vorticity of {\em one} general
KS state vanishes automatically
\begin{equation}
\mbox{\boldmath$\gamma$}_{KS}({\bf r})=
\mbox{\boldmath $\nabla$}
\times \mbox{\boldmath $\nabla$} \; \zeta({\bf r})=0
\label{vorticity-KS}
\end{equation}
irrespective of its special form.
If two electrons occupy the same orbital state then the density and 
paramagnetic current density have to be multiplied by the factor of 2 and
 {\em the vorticity of the whole KS system vanishes automatically}.

One could object that the above described procedure is not gauge invariant, 
because ${\bf j}^p({\bf r})$ is gauge dependent.
If we replace (\ref{det-j}) 
by equating the gauge invariant vorticities of both systems
and consider that the vorticity of the (singlet state in the) KS system vanishes 
\begin{equation}
\mbox{\boldmath$\gamma$}_{exact}({\bf r})\stackrel{!}{=}
\mbox{\boldmath$\gamma$}_{KS}({\bf r})=0
\label{det-vort}
\end{equation}
it is obvious that the exact vorticity has to vanish, 
what agrees with the result of the former approach.
However, in order to be absolutely correct we have 
to consider the following subtlety. 
Observe that the canonical orbital angular momentum 
$\sum_k {\bf r}_k \times {\bf p}_k$ is gauge dependent and it 
 is only conserved in the 
symmetric gauge (see Sect.IIA), which we used in out exact solutions. 
Therefore, we should characterize the only singlet state, 
which can be NIVR, in the following way: 
it is that state, which has $M_L=0$ in the symmetric gauge. 
This statement covers all gauges.


\newpage

\subsection{Triplet state: }
In a first step of sophistication we assume that 
 the KS system is circular and the 
KS states are eigen-functions of the orbital angular momentum
 $\varphi_k({\bf r})=e^{im_k \alpha} R_k(r)$.
(Do not mix up the relative angular momentum $m$ in in Sect.2
and the one-particle angular momentum $m_k$ in this section, which    
are both denoted by '$m$'.)
Therefore, we investigate, whether the exact system (which is circular
and conserves total orbital angular momentum) can be replaced by 
a circular KS system with the same total angular momentum.
This assumption is common practice in numerical  
self-consistent CDFT calculations and allows us to obtain  
the results in  Ref.\cite{Wensauer-Roessler1} as a special case. 
Equating both densities $(n,{\bf j}^p)$ provides the following 
set of equations
\begin{equation}
n_{KS}(r)=[R_1(r)]^2+[R_2(r)]^2 
\stackrel{!}{=} n_{exact}(r)
\label{det-n-trip}
\end{equation}
\begin{equation}
j^p_{KS}(r)=m_1\frac{[R_1(r)]^2}{r}+m_2\frac{[R_2(r)]^2}{r}
 \stackrel{!}{=} j^p_{ exact}(r)
\label{det-jp-trip}
\end{equation}
Now we investigate the limit for small $r$.
In the Appendix it is shown that 
for the special case of a circular KS system
$R_k(r) \rightarrow c_k \;r^{|m_k|}$.
Equations (\ref{n0}) and (\ref{jp0})
 can be written as $ n_{exact}(r) \rightarrow n_0+n_2 \;r^2$ and 
$\j^p_{exact}(r) \rightarrow j_0\; M_L\; r$.
Then (\ref{det-n-trip}) and (\ref{det-jp-trip}) 
read in the limit $r\rightarrow 0$ for the essential lowest terms 
\begin{equation}
c_1^2\; r^{2|m_1|}+c_2^2\; r^{2|m_2|}=n_0+n_2\; r^2
\label{n-r0}
\end{equation}
\begin{equation}
c_1^2\;m_1\; r^{2|m_1|-1}+c_2^2\;m_2\; r^{2|m_2|-1}=j_0\;M_L\;r
\label{jp-r0}
\end{equation}
(\ref{n-r0}) can be fulfilled only if one of the 
angular momenta vanishes and the modulus of the other one is unity, 
say $m_1=0$ and $|m_2|=1$. 
This choice satisfies also (\ref{jp-r0}).
Consequently, {\em only the exact states with $|M_L|=|m_1+m_2|=1$ 
can be NIVR by a circular KS system,}
 whereby only the state with $M_L=-1$ can be the ground state 
(see Fig.\ref{fig-s-t}).
In this case we obtain from (\ref{det-n-trip}) and (\ref{det-jp-trip})
explicit formulae for the radial parts
\begin{equation}
[R_2(r)]^2= -r\; j^p_{exact}(r)
\end{equation}
\begin{equation}
[R_1(r)]^2= n_{exact}(r)-[R_2(r)]^2
\end{equation}
agreeing with Ref.\cite{Wensauer-Roessler1}. 

In a second step, we abandon the constraint 
for circular symmetry of the KS system, 
because it cannot be taken for granted.
Additionally, we  equate the 
gauge invariant vorticities instead of the 
gauge-dependent paramagnetic current densities of the exact and the KS system.
The analytical properties of the KS functions for  $r \rightarrow 0$ 
 in this general case  are discussed in Appendix A.  
The gauge constants $m_{\cal G}$ can be neglected here 
because they cancel in gauge invariant quantities like 
$n({\bf r})$ and $\mbox{\boldmath$\gamma$}({\bf r})$ 
anyway.  If  we substitute 
$R_k(r) \rightarrow c_k\; r^{\widetilde{m}_k} $ 
in both  KS densities 
(\ref{det-n-trip}) and (\ref{det-jp-trip}),  and insert both
densities into the  the vorticity of the KS system 
\begin{equation} 
\mbox{\boldmath$\gamma$}_{KS}({\bf r})=\mbox{\boldmath $\nabla$}
\times \frac{{\bf j}^p_{KS}(r)} {n_{KS}(r) } \; {\bf e}_\alpha =
 \frac{1}{r}\frac{d}{dr}
\, \bigg( r\,\frac{j^p_{KS}(r)}{n_{KS}(r)} \bigg)\; {\bf e}_z
\equiv \gamma_{KS}(r) \;{\bf e}_z
\end{equation}
we obtain 
\begin{equation}
\gamma_{KS}(r) \rightarrow 
\frac{2\, c_1^2 \;c_2^2\; (\widetilde{m}_1-\widetilde{m}_2)^2\;
 r^{2(\widetilde{m}_1+\widetilde{m}_2-1)}}
{[c_1^2\; r^{2\widetilde{m}_1}+c_2^2\; r^{2 \widetilde{m}_2}]^2}
\end{equation}
The lowest-power term of the exact vorticity (\ref{gamma0}) can be  written as 
\begin{equation}
\gamma_{exact}(r) \rightarrow \gamma_0 \ M_L
\end{equation}
Consequently, the two constraints from equating $n(r)$ and $\gamma(r)$ read
\begin{equation}
c_1^2\; r^{2\widetilde{m}_1}+c_2^2\; r^{2\widetilde{m}_2}=n_0+n_2 \;r^2
\label{n-general}
\end{equation}
\begin{equation}
\frac{2\, c_1^2 \;c_2^2\; (\widetilde{m}_1-\widetilde{m}_2)^2\;
 r^{2(\widetilde{m}_1+\widetilde{m}_2-1)}}
{(n_0 +n_2\; r^2)^2}=\gamma_0\;M_L
\label{gamma-general}
\end{equation}
whereby in the denominator of the second constraint 
 the first constraint has been used.
Although (\ref{gamma-general}) and (\ref{jp-r0}) differ,
 the conclusion from (\ref{n-r0},\ref{jp-r0}) and 
(\ref{n-general},\ref{gamma-general}) are very similar.
(\ref{n-general},\ref{gamma-general}) {\em can be fulfilled only if
the small-r-exponents of the KS states are $0$ and $1$.}
The same conclusion can be drawn from equating the density and the 
paramagnetic current density for KS systems of arbitrary symmetry.


\newpage
\section{Summary and conclusions}
We used the exact solutions of the Schr\"odinger equation 
of a two-dimensional model system for the investigation of NIVR.
Qualitative conclusions can be drawn already on the basis of 
analytical results for the power expansion for $r\rightarrow 0$ 
 of the densities ${\cal N}({\bf r})$.

For {\em singlet} states ($M_L=even$) 
there is a simple and  straightforward proof 
without any assumptions that 
only the state with vanishing total orbital angular momentum $M_L=0$ 
can be NIVR.  This state is the ground state for small magnetic fields 
(see Fig.\ref{fig-s-t}) 
and it does not produce a paramagnetic current.

For {\em triplet} states ($M_L=odd$)
 a fully satisfactory statement can be made only 
under the assumption, that the KS system has the same circular symmetry 
as the real system  and consequently conserves orbital angular momentum.
Then only the exact states with $|M_L|=1$ are NIVR by KS states with 
angular momenta $m_1=0$ and $|m_2|=1$.  \\
If we allow KS systems with arbitrary symmetry, we can show that 
NIVR allows only KS states with small-r-exponents (see Appendix) 
of $0$ and $1$. However, we cannot say which {\em exact} 
states can or cannot be represented by symmetry-broken KS systems.
In other words, we know that the exact states with $|M_L|=1$ can be 
represented by  circular KS systems, and that all other states cannot be 
represented by  circular KS systems. 
This statement is valuable despite its limitations, because the assumption 
of circular symmetry in self-consistent calculations is common practice.

At the end we want to discuss the connection between NIVR and 
the property of being a ground state. 
The singlet states belonging to the full black line 
($M_L=0$) in Fig.\ref{fig-s-t} 
can be NIVR, no matter whether they are ground states for the 
given $\omega_0$ and $\omega_c$ or not.
All other states with even $M_L$ cannot, even if they are ground states.
The full blue line ($M_L=-1$), which is the ground state in the second 
$\omega_c$-region in Fig.\ref{fig-s-t},  can be NIVR everywhere.
Even the states belonging to the broken blue line ($M_L=+1$) are
NIVR, although they are never the ground state.
All other states with odd $M_L$ are not NIVR by a circular KS system.
Consequently, {\em  all those states at non-zero $B$ can be NIVR, 
which are 
 continuously connected to states at $B=0$, which
are the ground states for a given spin state at $B=0$,
no matter if they are the ground states at non-zero $B$ or not.}
The last statement sounds  similar to
the scaling property in  Ref.\cite{Wensauer-Roessler1}, 
but it is not identical (see Introduction).

\newpage
\appendix
\section{ General Kohn-Sham wave function for $\bf r\rightarrow 0$
and gauge dependence}
We are going to show that 
(apart from a gauge-invariant normalization factor)
 any KS wave function in the limit 
$r\rightarrow 0$ has the form 
\begin{equation}
\varphi_{KS}(r,\alpha) \rightarrow r^{\widetilde{m}} e^{i(\widetilde{m}+m_{\cal G})\alpha} 
\label{phi-KS-r0}
\end{equation}
where the {\em small-r-exponent} $\widetilde{m} \ge 0$ is a gauge-invariant integer, but the integer 
$m_{\cal G}$ depends on  the gauge. If the canonical angular momentum 
$l_z =(1/i) \partial / \partial \alpha$ is conserved, 
then $\widetilde{m}+m_{\cal G}$ is the 
canonical angular momentum $m$ of the KS state.
 Otherwise it is just a state and gauge-dependent integer. The 
limiting behavior of the density $n \rightarrow r^{2\widetilde{m}}$ 
is gauge-invariant, as expected, but the paramagnetic current density 
${\bf j}^p \rightarrow  (\widetilde{m}+m_{\cal G})\; 
r^{2\widetilde{m}-1}\, {\bf e}_\alpha$ 
depends on the gauge. Even if the canonical angular momentum is conserved, 
the exponent determining the radial WF (and the density) for small $r$ 
and the canonical angular momentum should not be mixed up.

The KS Hamiltonian has the general form
\begin{equation}
H_{KS}=\frac{1}{2} \biggl[\frac{1}{i}\mbox{\boldmath $\nabla$}+
{1\over c}{\bf A}({\bf r})\biggr]^2 + V({\bf r})
\label{H-KS}
\end{equation}
 The concrete form of the effective potentials in this equation 
and their connection to the 
XC-energy functional can be found in \cite{Vignale-Rasolt2}.
Now we apply a gauge transformation with $\chi=-c\, m_{\cal G} \,\alpha$ 
which changes the vector potential as follows
\begin{equation}
{\bf A} \rightarrow {\bf A}'={\bf A}+(\mbox{\boldmath $\nabla$}\chi )
={\bf A}-c\; m_{\cal G} \,\frac{1}{r} \,{\bf e}_\alpha
\label{gauge-trafo}
\end{equation}
This gauge transformation is equivalent to introducing
 a flux line with $m_{\cal G}$ flux quanta  at the center.
After rearranging terms and writing the Laplace operator in polar coordinates
we obtain
\begin{eqnarray}
H'&=&-\frac{1}{2} r^{-1/2}\frac{\partial^2} {\partial r^2} r^{1/2}
+\frac{1}{2\,r^2}
\biggl[\bigg( \frac{1}{i}\frac{\partial}{\partial\alpha}- m_{\cal G} \bigg)^2 
-\frac{1}{4}\biggr]                     \nonumber\\ 
&+& \frac{1}{c} m_{\cal G} \, \frac{\bf A}{r} \cdot {\bf e}_\alpha
+\frac{1}{2\,c}\frac{1}{i}
(\mbox{\boldmath $\nabla$}\cdot {\bf A})
+\frac{1}{c\,i}{\bf A}\cdot \mbox{\boldmath $\nabla$}
+\frac{1}{2\,c^2} {\bf A}^2
+V
\label{H-KS-polar}
\end{eqnarray}
Because in our model all external potentials are continuous, it 
is natural to assume that so is  $V$ and $\bf A$.
(Actually we have only to 
 assume that $V$ diverges weaker than $1/r^2$ and $\bf A$ 
weaker than $1/r$.) 
Then the eigenfunction in the limit $r \rightarrow 0$ 
is defined by the first line in (\ref{H-KS-polar}), which is independent 
of the specific form of $V$ and $\bf A$.
With the ansatz
 $$\varphi'=\frac{u(r)}{r^{1/2}}\; w(\alpha) $$
and after multiplication with $r^2 \,[1/u(r) w(\alpha)]\, r^{1/2}$
the variables in the  Schr\"odinger equation $H'\,\varphi'=\epsilon\; \varphi'$
 can be separated providing
\begin{equation}
\frac{1}{u(r)}\biggl[-\frac{1}{2} r^2 \frac{\partial^2} {\partial r^2}
- r^2 \epsilon \biggr] u(r) =
\frac{1}{2} \frac{1}{w(\alpha)} 
\biggl[-\bigg( \frac{1}{i}\frac{\partial}{\partial\alpha}- m_{\cal G} \bigg)^2
+\frac{1}{4}\biggr] w(\alpha) 
\label{SGL-sep}
\end{equation}
This equation can only be fulfilled if both sides are constant.
Using the technique of separation of variables, we can easily show 
that (\ref{phi-KS-r0}) is the solution, 
whereby the term with $r^2\, \epsilon$ on the l.h.s. 
can be neglected because we are interested in the limit $r\rightarrow 0$.

{\bf Acknowledgement}

This work was supported by the German Research Foundation (DFG)
in the Priority Program SPP 1145. We are indebted to K.Capelle and 
for valuable discussions and critically reading and improving 
the manuscript.

\end{document}